\documentclass[12pt]{article}
\newlength{\vshift}
\newlength{\hshift}

\usepackage{amssymb}
\usepackage{amsmath}
\usepackage{amsthm}
\usepackage{amsopn}
\def\beq{\begin{equation}}
\def\eeq{\end{equation}}
\def\bea{\begin{eqnarray}}
\def\eea{\end{eqnarray}}
\def\no{\nonumber}
\def\p{p\!\!\!/}
\def\D{D\!\!\!/}
\def\A{A\!\!\!/}
\begin{document}

\vspace*{3cm}
\begin{center}
{\Large\bf{Hydrogen and muonic-Hydrogen Atomic Spectra in
Non-commutative Space-Time }}\vskip 4em{ {\bf M.
Haghighat}\footnote{mansour@cc.iut.ac.ir} and {\bf M.
Khorsandi}\footnote{m.khorsandi.sh@gmail.com }}\vskip
1em {\it Department of Physics, Isfahan University of Technology,\\
Isfahan 84156-83111, Iran\\}
\end{center}
\vspace*{1.5cm}
\begin{abstract}
Comparing electronic Hydrogen  with muonic Hydrogen shows that the
discrepancy in measurement of the Lamb shift in the both systems are
relatively of order of $(\frac{m_\mu}{m_e})^{4-5}$.  We explore the
spectrum of Hydrogen atom in noncommutative $QED$ to compare the
noncommutative effects on the both bound states.  We show that in
the Lorentz violating noncommutative QED the ratio of NC-corrections
is $(\frac{m_\mu}{m_e})^3$ while in the Lorentz conserving NCQED is
$(\frac{m_\mu}{m_e})^5$.  An uncertainty about $1 \,Hz\ll 3\,kHz$ in
the Lamb shift of Hydrogen atom leads to an NC correction about $10
\,MHz$ in the Lorentz violating noncommutative QED and about $400
\,GHz$ in the Lorentz conserving noncommutative QED.

\end{abstract}
\section{Introduction}
As a simple system, hydrogen atom  with high precision measurements
in atomic transitions is one of the best laboratories to test QED
and new physics as well. Meanwhile, there is some discrepancy
between the recent measurement of the muonic hydrogen Lamb shift and
the corresponding proton radius and the CODATA value which is
obtained from the spectroscopy of atomic hydrogen and
electron-proton scattering \cite{codata}. There are two
possibilities to explain the discrepancy: 1- the theoretical
calculation within the standard model is incomplete. 2- existence of
new physics beyond the standard model. There are attempts to explain
the new physics by considering a new particle in MeV range where
many stringent limits suppress its existence \cite{MeV}. However,
the effective new interactions do not need necessarily new particles
to mediate the new interactions \cite{newint}. For instance,
noncommutative (NC) space can induce new interactions in QED without
adding new particles in the theory. Theoretical aspects of the
noncommutative space have been extensively studied by many
physicists \cite{theo}. Meanwhile, noncommutative standard model
(NCSM) via two different approaches is introduced in
\cite{calmet,melic1,chaichian3} and its phenomenological aspects are
explored in \cite{martin}. Here we would like to study two body
bound state in noncommutative space to explore the differences in
the electronic and muonic hydrogen spectrum.  There are many studies
on the hydrogen atom in the NC space-time
\cite{ho,chaichian1,chaichian2,stern1,stern2,moumni,saha,bertolami,bertolami1,banerjee,Balachandran}.
 However, the effect of NC-space on the hydrogen atom at the lowest order is
doubtful.  P. M. Ho and H. C. Kao \cite{ho} have shown that there is
not any correction on the space-space NC-parameter in this system.
Since all fields live on the same noncommutative space the
noncommutativity of a particle not only should be opposite to its
anti-particle  but also the NC-parameter of a charged particle
should be opposite to any other particle of opposite charge. In
fact, for the proton as a point particle $\theta_p=-\theta_e$ and
the corrections on the Coulomb potential coming from both particles
cancel out each other.  However, the proton is not a point particle
and the parameter of noncommutativity is an effective parameter and
is not equal to $-\theta_e$ \cite{chaichian2}. Furthermore, even if
the proton can be considered as a point particle, the space time
noncommutativity has some impact on the spectrum of the atom.
Meanwhile, in the Lorentz conserving NCQED the NC-parameter appears
as $\theta^2$ which is equal for a point particle and its
antiparticle. Therefore, it is reasonable to examine the hydrogen
atom in the NC space.  As $\theta_{\mu\nu}$ has dimension $-2$ one
expects an energy shift proportional to $(\theta
m^2)^2mc^2\alpha^2$. In fact, in the NC-space a larger energy shift
for the muonic Hydrogen is expected in comparison with the ordinary
Hydrogen which is in agreement with the experimental data.

In Section II we explore two-body bound state in NC-space.  In
section III we examine the $1S$-$2S$ transition and the Lamb-shift
for Hydrogen and muonic-Hydrogen and the $g$-factor for electron and
muon.  In section IV we give a brief review on the Lorentz
conserving NCQED and calculate the $g$-factor for electron and muon,
the $1S$-$2S$ transition and the Lamb-shift for Hydrogen and
muonic-Hydrogen.  In section V we summarize our results.

\section{2-body bound state in Noncommutative Space-Time}
Since the NC-parameters for a point particle and its antiparticle
 are opposite, the NC-correction on the potential for the space-space
 part of NC-parameter is zero at the lowest order.  It can be
   shown that the Coulomb potential in the Schrodinger equation is
proportional to $(\theta_p+\theta_e)_{ij}$ that is zero in the point
particle limit of the proton \cite{Stefano, ho}.  In the both
references the starting point is the Schrodinger equation in the NC
space. However, one can show how the NC field theory through the
Bethe-Salpeter (BS)-equation leads to the Schrodinger equation with
a modified potential\cite{2-body}.  In fact, it is better to start
from the NC-field theory (NCFT) to avoid the mistake on the
NC-contributions from each particle in the bound state.  For
instance, the correct potential in the NC-Schrodinger equation can
be explored in studying the kernel in the BS-equation for the
corresponding NCFT.  For this purpose examining the electron-proton
scattering amplitude in the NCQED is adequate to derive the
appropriate potential for the NCQM.

In a canonical noncommutative space, space-time coordinates are not
numbers but operators which do not commute
\begin{equation}\label{103}
[\hat{x}^{\mu},\hat{x}^{\nu}]=i\theta^{\mu\nu}=i\frac{C^{\mu\nu}}{\Lambda_{\tiny{NC}}^2},
\end{equation}
where $\theta^{\mu\nu}$ is the parameter of noncommutativity,
$C^{\mu\nu}$ is a constant and dimensionless antisymmetric tensor
and $\Lambda_{NC}$ is the noncommutative scale.  Since
noncommutative parameter, $\theta^{\mu\nu}$, is constant and
identifies  a preferred direction in  space, canonical version of
non-commutative space-time leads to the Lorentz symmetry violation.
There are two versions to construct the NCQED
\cite{calmet,melic1,chaichian3}.  In the first one in contrast with
the ordinary QED a momentum dependent phase factor appears in the
charged fermion-photon vertex as follows \cite{riad}
\begin{equation}\label{1}
ieQ\gamma_\mu \exp{(ip_\mu\theta^{\mu\nu}p'_\nu)},
\end{equation}
where $p_\mu$ and $p'_\nu$ are the incoming and outgoing momenta and
$\theta$ is the NC-parameter. Therefore, the electron-proton
amplitude can be written as
\begin{equation}\label{2}
{\cal{M}}_{NC}={\cal{M}}\exp({ip_\mu(\theta_e)^{\mu\nu}p'_\nu})\exp({ik_\mu(\theta_p)^{\mu\nu}k'_\nu}),
\end{equation}
where $p$ ($p'$) and $k$ ($k'$) are the incoming (outgoing) momenta
of the electron and proton, respectively. One can easily see that
the exponent in terms of the momentum transfer $q$ in the center of
mass is
\begin{equation}\label{3}
-i(\vec{\theta}_e+\vec{\theta}_p).\vec{p}\times\vec{q}
-ip_i(\theta_e^{i0}+\theta_p^{i0})q_0-i(p_0\theta_e^{0i}-k_0\theta_p^{0i})q_i,
\end{equation}
 where for $\theta_p=-\theta_e=\theta$ results in
\begin{equation}\label{4}
i(\sqrt{m_e^2+\bf{p}^2}+\sqrt{m_p^2+\bf{p}^2})\theta^{0i}q_i.
\end{equation}
Therefore, for $m^2_e$ and $m^2_p\gg\bf{p}^2$, the non-relativistic
potential in the NCQM is the Fourier transform of
\begin{equation}\label{5}
\frac{e^2\exp[i(m_e+m_p)\theta^{0i}q_i]}{q^2},
\end{equation}
 that leads to
$V_{NC}=V(\vec{r}-(m_e+m_p)\vec{\theta}_t)$ where
$\vec{\theta}_t=(\theta_{10}, \theta_{20}, \theta_{30})$. In fact,
to the lowest order of both $\theta$ and $\alpha$ the hydrogen atom
only receives some contribution from the temporal part of the
NC-parameter. Meanwhile, in \cite{stern1} and \cite{stern2} a new
correction due to the non commutativity of the source at the lowest
order of the space part of the NC-parameter is found.  Nonetheless,
it is of the order $\alpha^6$ and is not at the lowest order of
$\alpha$ too.

In the second approach, via Seiberg-Witten maps, the fields also
depend on the noncommutative parameter. Using Seiberg-Witten maps,
noncommutative standard model  and Feynman rules are fully provided
in references \cite{calmet,melic1}. In this approach, two
fermion-photon vertex is \cite{melic1}
\begin{equation}\label{104}
ieQ_{f}\gamma_{\mu}+\frac{1}{2}eQ_{f}[(p_{out}\theta
p_{in})\gamma_{\mu}-(p_{out}\theta)({\p}_{in}-m_{f})-({\p}_{out}-m_{f})(\theta
p_{in})_{\mu}] ,
\end{equation}
Where $Q$ is the fermion charge and $p_{in}$ and $p_{out}$ are
incoming and outgoing momenta, respectively. Considering proton as a
point particle, scattering amplitude of electron-proton in the
on-shell limit can be given as follows
\begin{equation}\label{105}
iM=\bar{u}(p')[-ie\gamma^{\mu}-\frac{1}{2}e(p'\theta_{e}
p)\gamma^{\mu}]u(p)(- ig_{\mu\nu}/
q^{2})\bar{\nu}(k)[ie\gamma^{\mu}+\frac{1}{2}e(k'\theta_{p}
k)\gamma^{\mu}]\nu(k'),
\end{equation}
where $p$ ($k$) and $p'$ ($k'$)  are incoming and outgoing momenta
of electron (proton), respectively.  At the lowest order of
$\theta_{\mu\nu}$, (\ref{105}) is the same as (\ref{2}) that means
in the second approach one has the same result as is given in
(\ref{5}).  In fact, for point particles in the QED bound states
such as positronium, there is no NC-correction at the lowest order
of $\alpha$ and $\theta_{ij}$.
\section{Hydrogen Atom in Noncommutative Space-Time }
In a two body bound state the point particles satisfy
$\theta_p=-\theta_e$ and the interaction potential only depends on
the time part of the NC-parameter as is shown in (\ref{5}).
Nevertheless, proton in the Hydrogen atom is not a point particle
and has an effective NC-parameter in terms of the NC-parameters of
its contents \cite{chaichian2} which is not equal to $-\theta_e$. In
fact, for the NC-parameter of the order of $1 TeV$ either the
electron with energy of order $eV$ or muon with $keV$-energy in the
Hydrogen cannot probe inside the proton to see its noncommutativity.
  It should be noted that for the muon the Bohr radius is about
$\frac{m_e}{m_{\mu}}a_e\sim 2.5 \times 10^{-13} m\sim 3r_p$ which is
comparable with the size of the proton. Meanwhile, for instance in
the Lamb shift of muon-Hydrogen the hadronic interactions are of the
order of $\alpha^2$ smaller than the QED interaction of the muon
with proton as a point particle. Thus, considering the NC-effects on
the muon interaction with the proton contents leads to corrections
of the order $m_{\mu}^2\theta\alpha^2$ smaller than what one finds
in the main part of the interactions. Therefore, even in the
muon-Hydrogen the NC-corrections due to the finite size of the
proton is negligible. In fact, at the low energy limit from the
noncommutative point of view proton is a macroscopic particle and
$\theta_{proton}\simeq 0$. Therefore, in the non-relativistic limit
and for the energy scale of atom, equation (\ref{3}) leads to
\begin{equation}\label{33}
-i\vec{\theta}_e.\vec{p}\times\vec{q}
-ip_i\theta_e^{i0}q_0-ip_0\theta_e^{0i}q_i.
\end{equation}
Equation (\ref{33}) for $\theta^{0i}= 0$ and $\theta^{ij}\neq 0$,
has been already considered to find bound on the NC-parameter in
Hydrogen atom \cite{chaichian1,stern1,stern2,moumni}.  The temporal
part of noncommutativity has some problem with the unitarity.
However, in some cases it can be shown that the quantum mechanics is
unitary for the temporal part of NC-parameter
\cite{Balachandran1,Balachandran2}.  Nevertheless, we examine the
temporal part ($\theta^{0i}\neq 0$ and $\theta^{ij}= 0$) in the
non-relativistic limit where (\ref{33}) leads to
\begin{equation}\label{poten1}
\frac{e^2\exp[im_e\theta^{0i}q_i]}{q^2},
\end{equation}
or
\begin{equation}\label{1061}
V_{nc}(\vec{r})=V(\vec{r}-m_e\vec{\theta}_t)-V(\vec{r})\simeq -
\alpha
m_{e}\frac{\overrightarrow{\theta}.\overrightarrow{r}}{r^{3}}.
\end{equation}
In (\ref{1061}) only the parallel part of $\vec{\theta}$ with
$\vec{r}$ has some contribution on the NC-potential that is
\begin{equation}\label{1071}
V_{nc}(r)= -m_{e} \alpha \frac{|\vec{\theta}|\cos{\phi}}{r^{2}},
\end{equation}
where $\cos{\phi}$ is the angle between $\vec{\theta}$ and $\vec{r}$
 and $\alpha $ is the fine structure constant. Using the
perturbation theory for the ground state leads to zero energy shift
for this state.  However, for the excited states in the
non-relativistic limit, the states $^2S_{\frac{1}{2}}$ and
$^2P_{\frac{1}{2}}$ are degenerate.  Therefore, the potential
(\ref{1071}) splits these states into 2 states with an energy
difference proportional to $<R_{n0}\mid\frac{-m_e\alpha
|\vec{\theta}|}{r^2}\mid R_{n1}>\sim -m^3_e\alpha^3 |\vec{\theta}|$.
But one can surprisingly show that
\begin{equation}\label{108'}
<R_{20}\mid\frac{1}{r^2}\mid R_{21}>=0,
\end{equation}
or up to the first order of $|\vec{\theta}|$ the states
$2^2S_{\frac{1}{2}}$ and $2^2P_{\frac{1}{2}}$ remain degenerate.
Nonetheless,  for $n=3$ one has
\begin{equation}\label{108}
\Delta E_{NC}^{H-atom}=\frac{2\sqrt{2}}{3}m^3_e\alpha^3
|\vec{\theta}|.
\end{equation}
Here we consider the effects of the NC-space-time on the physical
quantities such as $1S-3S$ transition, Lamb-shift in atom and the
anomalous magnetic moment to fix the NC-parameter in accordance with
the experimental uncertainties.

 {\bf $1S-3S$ Transition:}
  The obtained energy shift (\ref{108}), from the temporal part of noncommutativity, leads
  to an additional contribution on the theoretical
  value of $1S$-$3S$ transition in Hydrogen atom as follows
\begin{equation}\label{108'}
\Delta E_{NC}^{1S-3S}=\frac{2\sqrt{2}}{3}m^3_e\alpha^3
|\vec{\theta}|.
\end{equation}
 The uncertainty on the experimental value for $1S$-$3S$ transition
 in Hydrogen atom is about $13\hspace{.2cm} kHz $\cite{1s3s}
\begin{equation}\label{109}
f_{1S-3S}=2 922 742 936.729 (13) MHz.
\end{equation}
Therefore, for $\Lambda=1.5\hspace{.2cm}TeV$ one has
\begin{equation}\label{110}
\Delta E_{NC}^{1S-3S}\sim 30\hspace{.2cm} Hz\ll 13\hspace{.2cm} kHz.
\end{equation}

{\bf Lamb Shift:} For $\Lambda=1.5\hspace{.2cm}TeV$ the
NC-correction on the lamb shift for the electronic hydrogen is
\begin{equation}\label{lamb-e}
\Delta E_{NC}^{H_e}=\frac{2\sqrt{2}}{3}m^3_e\alpha^3
|\vec{\theta}|\sim 30\hspace{.2cm} Hz\ll 48\hspace{.2cm} kHz,
\end{equation}
where $48\hspace{.2cm} kHz$ is the experimental accuracy on the
$n=3$ lamb shifts in the Hydrogen atom\cite{lamb3s}. Meanwhile, for
the muonic hydrogen one has
\begin{equation}\label{lamb-mu}
\Delta E_{NC}^{H_{\mu}}=(\frac{m_{\mu}}{m_e})^{3}\Delta
E_{NC}^{H_e}= 240\hspace{.2cm} MHz\sim 2\times10^{-4} meV.
\end{equation}

{\bf g-2 for electron and muon:} Since the NC-parameter has
dimension -2 the dimensionless quantity $a=\frac{g-2}{2}$ should be
corrected, at the lowest order in NC-space, as
$C(\frac{\alpha}{2\pi})p_\mu\theta^{\mu\nu}p'_\nu$, where $C$ is a
constant which is obtained in \cite{mortazavi,riad}.  Therefore, for
$\Lambda=1.5\hspace{.2cm} TeV$ the NC-correction on $a$ for electron
is
\begin{equation}\label{lamb-e}
\delta a_e=\frac{5}{6}\frac{\alpha}{2\pi}\frac{p^2}{\Lambda^2}
\simeq\frac{5}{6}\frac{\alpha}{2\pi}\frac{m_e^2}{\Lambda^2}\sim10^{-16},
\end{equation}
and for the muon where in E286 experiment has a momentum about
$3\hspace{.2cm} GeV$ is
\begin{equation}\label{lamb-e}
\delta a_\mu=\frac{5}{6}\frac{\alpha}{2\pi}\frac{p^2}{\Lambda^2}
\simeq\frac{5}{6}\frac{\alpha}{2\pi}(\frac{3}{1500})^2\sim
3\times10^{-9}.
\end{equation}

\section{Hydrogen atom in Lorentz Conserving Noncommutative Space-Time}
As the NC-parameter is a real and constant Lorentz tensor, there is,
obviously, a preferred direction in a given particle Lorentz frame
which leads to the Lorentz symmetry violation. On the other hand,
experimental inspections for Lorentz violation, including clock
comparison tests, polarization measurements on the light from
distant galaxies, analyses of the radiation emitted by energetic
astrophysical sources, studies of matter-antimatter asymmetries for
trapped charged particles and bound state systems \cite{LV} and so
on, have thus far failed to produce any positive results. These
experiments strictly bound the Lorentz-violating parameters,
therefore, in the lower energy limit, the Lorentz symmetry is an
almost exact symmetry of the nature \cite{wolf}. However, Carlson,
Carone, and Zobin (CCZ) have constructed Lorentz-conserving
noncommutative quantum electrodynamics based on a contracted Snyder
algebra \cite{Carlson}. In this class of NC theories, the parameter
of noncommutativity is not a constant but an operator which
transforms as a Lorentz tensor. In fact, (\ref{103}) should be
extended to
\begin{eqnarray}\label{103'}
[\hat{x}^{\mu},\hat{x}^{\nu}]=i\hat{\theta}^{\mu\nu},\,\,\,
[\hat{\theta}^{\alpha\beta},\hat{\theta}^{\mu\nu}]=0,\,\,\,[\hat{\theta}^{\mu\nu},\hat{x}^{\nu}]=0,
\end{eqnarray}

where $\hat{\theta}^{\mu\nu}$ is an operator.  Consequently,
according to the Weyl- Moyal correspondence, to construct the LCNC
action the ordinary product should be replaced with the star product
as follows
\begin{equation}\label{starpro}
f\ast g(x,\hat{\theta)}=
f(x,\theta)exp(i/2\overleftarrow{\partial}_{\mu}\theta^{\mu\nu}\overrightarrow{\partial}_{\nu})g(x,\theta).
\end{equation}
In this formalism a sufficiently fast falling weight function
$W(\theta)$ has been used to construct the Lorentz invariant
lagrangian in a non-commutative space as
\begin{equation}\label{110'}
\mathcal{L}(x)=\int d^{6}\theta W(\theta)
\mathcal{L}(\phi,\partial\phi)_{\ast},
\end{equation}
where the Lorentz  invariant  weight  function $W(\theta)$ is
introduced to suppresses the NC-cross section for energies beyond
the NC-energy scale.  In reference \cite{exist} on the existence of
an invariant normalized weight function is discussed and an explicit
form for $W(\theta)$ is given in terms of Lorentz invariant
combinations of $\theta^{\mu\nu}$'s.  The function $W(\theta)$ can
be used to define an operator trace as
\begin{equation}\label{106}
Tr\hat{f}=\int d^{4}x d^{6}\theta W(\theta)f(x,\theta),
\end{equation}
in which $W(\theta)$ has the following properties
\begin{equation}
\int d^{6}\theta W(\theta) = 1,
\end{equation}

\begin{equation}\label{107}
\int d^{6}\theta W(\theta)\theta^{\mu\nu}=0,
\end{equation}

\begin{equation}\label{1082}
\int d^{6}\theta W(\theta)\theta^{\mu\nu}\theta^{\kappa\lambda}=
\langle\frac{\theta ^{2}}{2}\rangle
(g^{\mu\nu}g^{\mu\lambda}-g^{\mu\lambda}g^{\nu\kappa}),
\end{equation}
where
\begin{equation}\label{109'}
\langle\theta^{2}\rangle=\int d^{6}\theta
W(\theta)\theta^{\mu\nu}\theta_{\mu\nu}.
\end{equation}
As (\ref{107}) shows in the expansion of the Lagrangian (\ref{110'})
in terms of the NC-parameter the odd powers of $\theta_{\mu\nu}$
vanishes.  In fact, to obtain the nonvanishing $\theta$-dependence
terms, all fields should be expanded at least up to the  second
order of the NC-parameter.  The Lorentz conserving NCSM is fully
introduced in \cite{haghighat} and its fermionic part where we are
interested to explore is given as follows

\begin{eqnarray}\label{111}
S_{fermion}= \int d^{4}x (\overline{L}i{\D}L + \overline{R}i{\D}R)+
\int d^{6}\theta \int d^{4}x
W(\theta)\theta^{\mu\nu}\theta^{\kappa\lambda}
(-\frac{i}{8}\overline{L} \gamma^{\rho}F^{0}_{\mu\kappa}F^{0}_{\lambda\rho}D^{0}_{\nu}L\no\\
-\frac{i}{4}\overline{L}\gamma^{\rho}F^{0}_{\mu\rho}F^{0}_{\nu\kappa}D^{0}_{\lambda}L
-\frac{1}{8}\overline{L}\gamma^{\rho}(D^{0}_{\mu}F^{0}_{\kappa\rho})D^{0}_{\nu}D^{0}_{\lambda}L
-\frac{i}{8}\overline{L} \gamma^{\rho}F^{0}_{\mu\nu}F^{0}_{\kappa\rho}D^{0}_{\lambda}L)\hspace{1cm} \no\\
\int d^{6}\theta \int d^{4}x
W(\theta)\theta^{\mu\nu}\theta^{\kappa\lambda}
(-\frac{i}{8}\overline{R}
\gamma^{\rho}F^{0}_{\mu\kappa}F^{0}_{\lambda\rho}D^{0}_{\nu}R
-\frac{i}{4}\overline{R}\gamma^{\rho}F^{0}_{\mu\rho}F^{0}_{\nu\kappa}D^{0}_{\lambda}R\hspace{1cm}\no\\
-\frac{1}{8}\overline{R}\gamma^{\rho}(D^{0}_{\mu}F^{0}_{\kappa\rho})D^{0}_{\nu}D^{0}_{\lambda}R
-\frac{i}{8}\overline{R}
\gamma^{\rho}F^{0}_{\mu\nu}F^{0}_{\kappa\rho}D^{0}_{\lambda}R),\hspace{3cm}
\end{eqnarray}
where $L$ and $R$ stand, respectively, for left and right handed
fermions and $F^{0}_{\mu\nu}$ is the ordinary field strength in the
standard model.  To find the LCNC-effects, at the lowest order, on
the hydrogen atom we only consider the QED part of the NC-action
(\ref{111}) as follows
\begin{eqnarray}\label{112}
\int d^{6}\theta \int d^{4}x
eQ_{f}(\overline{\nu}_{L}{\A}_{0}\nu_{L}-\frac{1}{8}\theta^{\mu\nu}\theta^{\kappa\lambda}\overline{\nu}_{L}\gamma^{\rho}
\partial_{\mu}A_{0\kappa\rho}\partial_{\nu}\partial_{\lambda}\nu_{L}+\overline{e}{\A}_{0}e_{L}-\no\\
\frac{1}{8}\theta^{\mu\nu}\theta^{\kappa\lambda}\overline{e}\gamma^{\rho}\partial_{\mu}A_{0\kappa\rho}\partial_{\nu}\partial_{\lambda}e),\hspace{3cm}
\end{eqnarray}
Where $A_{0\mu\nu} = \partial_{\mu}A_{0\nu}-\partial_{\nu}A_{0\mu}$
 and the charged fermions interact with photon via the following
 vertex
\begin{equation}\label{113}
ieQ_{f}\gamma^{\mu}(1+
\frac{\langle\theta^{2}\rangle}{96}(\frac{q^{4}}{4}-m_{f}^{2}q^{2})
).
\end{equation}
In (\ref{113}) the $\theta$-dependence is appeared as
$\langle\theta^{2}\rangle$ which is similar for both particle and
its antiparticle.  In fact, in LCNC-QED in contrast with NCQED
according to $\theta_{f_-}= -\theta_{f_+}$, particle vertex doesn't
cancel antiparticle vertex.  Therefore, in $f_-f_+$ bound state the
LCNC effect via the $f_-f_+$ scattering amplitude, at low energy
limit, leads to a potential in momentum space as
\begin{equation}\label{114}
\widetilde{V}(q)=
-\frac{e^{2}}{q^{2}}-\frac{e^{2}(m_{f^-}^{2}+m_{f^+}^{2})\langle
\theta^{2}\rangle}{96},
\end{equation}
Where to obtain (\ref{114}), at low momentum transfer, the second
term in (\ref{113}) is ignored in comparison with the third one.
However, proton as a particle with internal structure doesn't see
the NC-space in those systems which the momentum transfer is small
such as Hydrogen like atom. Therefore, the NC-potential in the
Hydrogen atom is

\begin{equation}\label{hyd}
\widetilde{V}(q)= -\frac{e^{2}}{q^{2}}-\frac{e^{2}m_{e^-}^{2}\langle
\theta^{2}\rangle}{96},
\end{equation}
or
\begin{equation}\label{115}
V(r) =- \frac{e^{2}}{4\pi r}-\frac{e^{2}m_{e}^{2}\langle
\theta^{2}\rangle}{96}\delta(r).
\end{equation}
The NC-correction on the Coulomb potential in (\ref{115}) is small
and its expectation value directly gives the energy shift on the
 energy levels of the Hydrogen atom as follows
\begin{equation}\label{117}
\Delta E_{LCNC}^{H-atom}= - \langle
\psi|\frac{e^{2}m_{e^-}^{2}\langle
\theta^{2}\rangle}{96}\delta(r)|\psi\rangle=
-\frac{e^{2}m_{e^-}^{2}\langle \theta^{2}\rangle}{96}
|\psi_{nl}(r=0)|^{2} ,
\end{equation}
Where $ |\psi_{nl}(r=0)|^{2}=\frac{\alpha^{3}m_{e^-}^{3}}{\pi
n^{3}}\delta_{l0}$ leads to
\begin{equation}\label{118}
\Delta E_{LCNC}^{H-atom}= -
\frac{m_{e^-}^{5}\alpha^{4}\langle\theta^{2}\rangle}{24n^{3}}
\delta_{l0},
\end{equation}
or with $\Lambda_{LCNC}=(\frac{12}{\langle\theta^{2}\rangle})^{4}$,
(\ref{118}) can be rewritten as
\begin{equation}\label{119} \Delta
E_{LCNC}^{H-atom}=
-\frac{m_{e^-}^{5}\alpha^{4}}{2n^{3}}\frac{1}{\Lambda_{LCNC}^{4}}\delta_{l0}.
\end{equation}
Here we fix the NC-parameter by the most precise experimental value
in the Hydrogen atom (i.e. $1S-2S$ transition) then we find the
effects of the NC-space-time on the other physical quantities such
as Lamb-shift in atom and the anomalous magnetic moment.

 {\bf
$1S-2S$ Transition:} The experimental value for $1S$-$2S$ transition
in the Hydrogen atom \cite{phil} can fix the upper bound on the
parameter $\Lambda_{LCNC}$ as follows

\begin{equation}\label{1s2s}
\Delta E_{LCNC}^{1S-2S}=
\frac{7m_{e}^{5}\alpha^{4}}{16}\frac{1}{\Lambda_{LCNC}^{4}}\sim
34\hspace{.2cm} Hz,
\end{equation}
leads to
\begin{equation}\label{119}
\Lambda_{LCNC}\sim 0.2\hspace{.2cm} GeV.
\end{equation}

{\bf Lamb Shift:} For $\Lambda=0.5\hspace{.2cm} GeV$ the
NC-correction on the lamb shift for electronic hydrogen is
\begin{equation}\label{lc-lamb-e}
\Delta
E_{NC}^{H_e}=\frac{m_{e}^{5}\alpha^{4}}{16}\frac{1}{\Lambda_{LCNC}^{4}}\sim
0.1\hspace{.2cm} Hz \ll 3\hspace{.2cm} kHz,
\end{equation}
where $48\hspace{.2cm} kHz$ is the experimental accuracy on the
$n=3$ lamb shifts in the Hydrogen atom\cite{eides}.  Meanwhile, for
the muonic hydrogen one has
\begin{equation}\label{lc-lamb-mu}
\Delta E_{NC}^{H_{\mu}}=(\frac{m_{\mu}}{m_e})^{5}\Delta
E_{NC}^{H_e}\simeq 40 \hspace{.2cm}GHz\sim 0.03\hspace{.2cm} meV.
\end{equation}

{\bf g-2 for electron and muon:} As Eq.(\ref{113}) shows the
NC-correction on $a=\frac{g-2}{2}$ should be proportional to $q^2$
which leads to zero NC-correction on $a$ at zero momentum transfer.

\section{summary}
In this paper two body bound state has been studied by examining the
scattering amplitude in the Lorenz violated (LV) and Lorentz
conserving (LC) NCQED as given in (\ref{2}) and (\ref{114}),
respectively. For a bound state of a particle with its antiparticle
the NC potential in LVNCQED, in contrast with LCNCQED, only depends
on the space-time part of the NC-parameter, see (\ref{3}).  As the
proton in Hydrogen atom is not a point particle,
$\theta_p\neq-\theta_e$. In fact, in $ep$-scattering in the low
energy limit, the electron cannot prob inside the proton to see its
NC-effects. In the Lorentz violated NCQED for Hydrogen
and muonic Hydrogen atom we have found:\\
1- In $1S-2S$ transition in the Hydrogen atom the NC-effect is zero, see(\ref{108'}).\\
2- In $1S-3S$ transition in the Hydrogen atom the NC-parameter of
the order of $\Lambda_{NC}=1.5\,\, TeV$ leads to a small
correction on the theoretical value which is not detectable.\\
3- $\Lambda_{NC}=1.5\,\, TeV$ leads to an NC-shift on the $n=3$ Lamb
shift about $30\,\,Hz$ which is far from the $48\,\,kHz$ current
uncertainty on the Lamb shift in the Hydrogen atom. As
(\ref{lamb-mu}) shows an uncertainty of order of $3\,\,kHz$ in the
Lamb shift of Hydrogen leads to
$(3\,kHz)(\frac{m_\mu}{m_e})^3=26\,\,GHz$ which can only explain a
small part of the current deviation between the experimental
measurement
 and the theoretical prediction. \\
4- The NC-effect on the g factors of electron and muon are
$a_e=10^{-16} $ and $a_\mu=10^{-9}$, respectively.  The obtained
values for $a_e$ and $a_\mu$ are in agreement with the experimental
measurements.\\
5-$\Lambda_{NC}=1.5\,\, TeV$ which is found from the muon g-factor
is a stringent bound in the atomic systems where the NC-parameter is
usually of the order of a few $GeV$ \cite{stern1,stern2}.  However,
this bound can not explain the discrepancy in measurement of Lamb
shift in muonic Hydrogen.\\

 In the Lorentz conserving NCQED for Hydrogen
and muonic Hydrogen atom we have found:\\
1- In $1S-2S$ transition in the Hydrogen atom the NC-parameter has
been fixed about $\Lambda_{NC}=0.5\,\, GeV$ which is the
first bound on this parameter in an atomic system.\\
2- $\Lambda_{NC}=0.5\,\, GeV$ leads to an NC-shift on the
$2s_{1/2}-2p_{1/2}$ transition in hydrogen atom about $0.1\,\,Hz$
which is far from the $3\,\,kHz$ current uncertainty on the Lamb
shift in the Hydrogen atom.  As (\ref{lc-lamb-mu}) shows an
uncertainty of order of $3\,\,Hz\ll 3\,\,kHz$ in the Lamb shift of
Hydrogen leads to $(3\,Hz)(\frac{m_\mu}{m_e})^5=1000\,\,GHz$ which
can explain the current difference between the experimental
measurement and
the theoretical prediction about $0.3\, meV$ \cite{codata}. \\
3- The NC-correction on $a=\frac{g-2}{2}$ is proportional to $q^2$
which is zero at the zero momentum transfer. \\



\begin{thebibliography}{99}
\bibitem{codata}
P. J. Mohr, B. N. Taylor, and D. B. Newell, Rev. Mod. Phys. {\bf 80}
, 633 (2008); R. Pohl et al. , Nature {\bf 466} , 213 (2010).
\bibitem{MeV}
V. Barger, Cheng-Wei Chiang, Wai-Yee Keung and D. Marfatia, Phys.
Rev. Lett. {\bf 106}, 153001 (2011); D. Tucker-Smith and I. Yavin,
Phys. Rev. D {\bf 83}, 101702 (2011); M. Williams, C.P. Burgess, A.
Maharana and F. Quevedo, JHEP {\bf 1108}, 106 (2011).
\bibitem{newint}  R. Onofrio, EPL, {\bf 104}, 20002 (2013);
 R. Onofrio, Mod. Phys. Lett. A {\bf 28}, 1350022 (2013); L.-B. Wang and
W.-T. Ni, Mod. Phys. Lett. A {\bf 28}, 1350094 (2013).
\bibitem{theo}
Everton M. C. Abreu, and M.J. Neves, Nucl. Phys. B {\bf 884}, 741
(2014);
 C.P. Martin, Phys. Rev. D {\bf 89}, 065018 (2014); Josip
Trampetic, and Jiangyang You, SIGMA {\bf 10}, 054 (2014); C.P.
Martin, Class. Quant. Grav. {\bf 30}, 155019 (2013); Elisabetta Di
Grezia, Giampiero Esposito, Marco Figliolia, and Patrizia Vitale,
Int. J. Geom. Meth. Mod. Phys. {\bf 10}, 1350023 (2013); Everton
M.C. Abreu, and Mario J. Neves, Int. J. Mod. Phys. A {\bf 28},
1350017 (2013); C.P. Martin, Phys. Rev. D {\bf 86}, 065010 (2012);
H. Falomir, F. Vega, J. Gamboa, F. Mendez, and M. Loewe, Phys. Rev.
D {\bf 86}, 105035 (2012); R. Horvat, A. Ilakovac, P. Schupp, J.
Trampetic, and J. You, JHEP {\bf 1204}, 108 (2012); Raul Horvat,
Amon Ilakovac, Peter Schupp, Josip Trampetic, and Jiang-Yang You,
Phys. Lett. B {\bf 715}, 340 (2012); Raul Horvat, and Josip
Trampetic, JHEP {\bf 1101}, 112 (2011);  M. Chaichian, P.
Presnajder, M. M. Sheikh-Jabbari, and A. Tureanu, Phys. Lett. B {\bf
683}, 55 (2010); M. M. Ettefaghi and M. Haghighat, Phys. Rev. D {\bf
77}, 056009 (2008); M. M. Ettefaghi, M. Haghighat, and R. Mohammadi,
Phys. Rev. D {\bf 82}, 105017 (2010); C.P. Martin, Phys. Rev. D {\bf
82}, 085020 (2010); L. Bonora and M. Salizzoni, Phys. Lett. {\bf
B504}, 80 (2001); C. P. Martin and D. Sanchez-Ruiz, Phys. Rev. Lett.
{\bf 83}, 476 (1999);
 M. M. Sheikh-Jabbari, JHEP  {\bf 9906}, 015 (1999);
 T. Krajewski and R. Wulkenhaar, Int. J. Mod. Phys.  {\bf A15}, 1011
 (2000); S. Minwalla, M. Van Raamsdonk and N. Seiberg, JHEP {\bf 0002}, 020
(2000); A. Matusis, L. Susskind and N. Toumbas, JHEP {\bf 0012}, 002
 (2000); M. Hayakawa, Phys.Lett. {\bf B478}, 394 (2000);
 A. A. Bichl, J. M. Grimstrup, L. Popp, M. Schweda and R. Wulkenhaar,
Int. J. Mod. Phys. {\bf A17} 2219 (2002); B. Jurco, S. Schraml, P.
Schupp and J. Wess, Eur. Phys. J. {\bf C17}, 521(2000); M. Buric, D.
Latas and V. Radovanovic, JHEP {\bf 0602}, 046 (2006); M. Buric, V.
Radovanovic and J. Trampetic, JHEP {\bf 0703}, 030 (2007); R.
Wulkenhaar,  JHEP {\bf 0203}, 024 (2002): R. Amorim, and Everton M.
C. Abreu, Phys. Rev. D {\bf 80}, 105010 (2009); R. Amorim1, Everton
M. C. Abreu, and W. G. Ramirez, Phys. Rev. D {\bf 81}, 105005
(2010).
\bibitem{calmet}
X. Calmet, B. Jurco, P. Scupp, J. Wess and M.wohlgenannt, Eur. Phys.
J. C {\bf 23}, 363 (2002).
\bibitem{melic1}
B. Meli$\acute{c}$, K. Passeka-Kumeri$\check{c}$ki, J.
Trampeti$\check{c}$, P. Schupp and M. Wohlgenannt, Eur. Phys. J C
{\bf 42}, 483 (2005); B. Meli$\acute{c}$, K.
Passeka-Kumeri$\check{c}$ki, J. Trampeti$\check{c}$, P. Schupp and
M. Wohlgenannt, Eur. Phys. J C {\bf 42}, 499 (2005).
\bibitem{chaichian3}
M. Chaichian, P. Presnajder, M. M. sheikh-Jabbari and A. Tureanu,
Eur. Phys. J. C {\bf 29}, 413(2003); M. Chaichian, P. Presnajder, M.
M. sheikh-Jabbari and A. Tureanu, Phys. Lett. B {\bf 526}, 132
(2002).
\bibitem{martin}
M.M. Ettefaghi, and R. Moazzemi, JCAP {\bf 1302}, 048 (2013);
Weijian Wang, Jia-Hui Huang, and Zheng-Mao Sheng, Phys.Rev. D {\bf
88}, 025031 (2013); A. Jafari, Eur. Phys. J. C {\bf 73}, 2271
(2013); S. Aghababaei, M. Haghighat, and A. Kheirandish, Phys. Rev.
D {\bf 87}, 047703, (2013); M. M. Ettefaghi, Phys. Rev. D {\bf 86},
085038 (2012); Weijian Wang, Jia-Hui Huang, and Zheng-Mao Sheng,
Phys. Rev. D {\bf 86}, 025003 (2012);  Weijian Wang, Feichao Tian
and Zheng-Mao Sheng, Phys. Rev. D {\bf 84},  045012 (2011) ; G.
Deshpande, and Sumit K. Garg, Phys. Lett. B {\bf 708}, 150 (2012);
Josip Trampetic, Int. J. Geom. Meth. Mod. Phys. {\bf 09}, 1261016
(2012); Horvat, A. Ilakovac, P. Schupp, J. Trampetic, and J. You,
JHEP {\bf 1204}, 108 (2012); Kai Ma, and Sayipjamal Dulat, Phys.
Rev. A {\bf 84}, 012104 (2011);  M. Haghighat, Phys. Rev. D {\bf 79
}, 025011 (2009); M. M. Ettefaghi , Phys Rev. D {\bf 79 }, 065022
(2009); R. Horvat and J. Trampetic, Phys. Rev. D {\bf 79 }, 087701
(2009); A. Joseph, Phys. Rev. D {\bf 79 }, 096004 (2009); M.
Haghighat, M. M. Ettefaghi, and M. Zeinali, Phys. Rev. D {\bf 73},
013007 (2006);  Mansour Haghighat, Nobuchika Okada and Allen Stern,
Phys. Rev. D {\bf 82}, 016007 (2010);  M. Zarei, E. Bavarsad, M.
Haghighat, I. Motie, R. Mohammadi and Z. Rezaei, Phys. Rev. D {\bf
81}, 084035 (2010);  C. P. Martin, D. Sanchez-Ruiz and C. Tamarit,
JHEP {\bf 0702}, 065 (2007); A. Al- boteanu, T. Ohl and R. Rckl,
Phys. Rev. D {\bf 74}, 096004 (2006); Seyed Yaser Ayazi, Sina
Esmaeili, Mojtaba Mohammadi-Najafabadi, Phys. Lett. B {\bf 712}, 93
(2012); V. Nazaryan and Carl E. Carlson, Int. J. Mod. Phys. A {\bf
20}, 3495 (2005); Amir H. Fatollahi and A. Jafari, Eur. Phys. J. C
{\bf 46}, 235 (2006); M. Haghighat, S. M. Zebarjad and F. Loran,
Phys. Rev. D. {\bf 66} , 016005 (2002);  A. Devoto, S. Di Chiara and
W. W. Repko, Phys. Rev. D {\bf 72}, 056006 (2005);  S. M. Carroll,
J. A. Harvey, V. A. Kostelecky, C. D. Lane, and T. Okamoto, Phys.
Rev. Lett. {\bf 87}, 141601 (2001); S. Godfrey and M. A. Doncheski,
Phys. Rev. D {\bf 65}, 015005 (2002).
\bibitem{ho}
P. M. Ho and H. C. Kao, Phys. Rev. Lett. {\bf 88},  151602 (2002).
\bibitem{chaichian1}
M. Chaichian, M. M. Sheikh-Jabbari and A. Tureanu, Phys. Rev. Lett.
{\bf 86}, 2716 (2001).
\bibitem{chaichian2}
M. Chaichian, M. M. Sheikh-Jabbari and A. Tureanu, Eur. Phys. J. C
{\bf} 36, 251 (2004).
\bibitem{stern1}
A. Stern, Phys. Rev. Lett. {\bf 100}, 061601 (2008).
\bibitem{stern2}
A. Stern, Phys. Rev. D {\bf 78}, 065006 (2008).
\bibitem{moumni}
M. Moumni and A. BenSlama, Int. J. Mod. Phys. Lett. A {\bf 28},
1350139 (2013).
\bibitem{saha}
A. Saha, Eur. Phys. J. C {\bf 51}, 199 (2007).
\bibitem{bertolami}
O. Bertolami, J. G. Rosa, C. M. L. de Arag$\tilde{a}$o, P.
Castorina, and D. Zappal$\grave{a}$, Phys. Rev. D {\bf 72}, 025010
(2005).
\bibitem{bertolami1}
O. Bertolami, J. G. Rosa, J. Phys.: Conf. Ser. {\bf 33}, 118 (2006).
\bibitem{banerjee}
R. Banerjee, B. Dutta Roy, S. Samanta, Phys. Rev. D {\bf 74}, 045015
(2006).
\bibitem{Balachandran}
E. Akofor, A.P. Balachandran, A. Joseph, L. Pekowsky, and B.A.
Qureshi, Phys. Rev. D {\bf 79}, 063004 (2009).
\bibitem{Stefano}
 S. Bellucci and A. Yeranyan, Phys. Lett. B {\bf 609}, 418 (2005).
 \bibitem{2-body}
 M. Haghighat and F. Loran, Phys. Rev. D.
{\bf 67},  096003 (2003); M. Haghighat and F. Loran, Mod. Phys.
Lett. A {\bf 16}, 1435 (2001).
\bibitem{riad}
I. F. Riad and M. M. Sheikh-Jabbari, J. High Energy Phys. {\bf
0008}, 045 (2000).
\bibitem{Balachandran1}
A. P. Balachandran, T. R. Govindarajan, C. Molina and P.
Teotonio-Sobrinho, JHEP {\bf 0410}, 072 (2004).
\bibitem{Balachandran2}
A. P. Balachandran and A. Pinzul, Mod. Phys. Lett. A {\bf 20}, 2023
(2005).
\bibitem{1s3s}
 O. Arnoult,
F. Nez, L. Julien, and F. Biraben ,  Eur. Phys. J. D {\bf 60}, 243
(2010).
\bibitem{lamb3s}
 C. W. Fabjan and F. M. Pipkin, Phys. Rev. A {bf 6}, 556 (1972).
\bibitem{mortazavi}
M. Haghighat and N. Mortazavi, IJPR. {\bf 11}, No.3, 265(2011).
\bibitem{LV}
S. M. Carroll, J. A. Harvey, V. A. Kostelecky, C.D. Lane, and T.
Okamoto, Phys. Rev. Lett. {\bf 87}, 141601 (2001); V. A. Kostelecky
and M. Mewes, Phys. Rev. Lett. {\bf 87}, 251304 (2001); T. Jacobson,
S. Liberati, and D. Mattingly, Nature {\bf 424}, 1019 (2003); B.
Altschul, Phys. Rev. Lett. {\bf 96}, 201101 (2006); Phys. Rev. D
{\bf 75}, 023001 (2007); G. Gabrielse, A. Khabbaz, D. S. Hall, C.
Heimann, H. Kalinowsky, and W. Jhe, Phys. Rev. Lett. {\bf 82}, 3198
(1999); D. F. Phillips, M. A. Humphrey, E. M. Mattison, R. E.
Stoner, R. F. C. Vessot, and R. L. Walsworth, Phys. Rev. D {\bf 63},
111101 (2001); P. Wolf, F. Chapelet, S. Bize, and A. Clairon, Phys.
Rev. Lett. {\bf 96}, 060801 (2006).
\bibitem{wolf}
P. Wolf, F. Chapelet, S. Bize, and A. Clairon, Phys. Rev. Lett. {\bf
96}, 060801 (2006); V. A.  Kostelecky and  M. M EWES, Phys. Rev.
Lett. {\bf 87}, 251304 (2001); T. Jacob- Son  S.  Liberati,  and  D.
Mattingly, Nature {\bf 424}, 1019 (2003); B.  Altschul,  Phys. Rev.
Lett. {\bf 96}, 201101  (2006).
\bibitem{Carlson}
C. E. Carlson, C. D. Carone, and N. Zobin,
 Phys. Rev. D {\bf 66}, 075001 (2002).
\bibitem{exist}
 Christopher D. Carone and Herry J. Kwee, Phys. Rev. D {\bf 73}, 096005 (2006).
\bibitem{haghighat}
M.  M. Ettefaghi and M. Haghighat,  Phys.  Rev.  D {\bf 75},
125002(2007); M. Haghighat and M. M. Ettefaghi, Phys. Rev. D {\bf
70}, 034017 (2004).
\bibitem{phil}
T.W. H$\ddot{a}$nsch et al,  Phil. Trans. R. Soc. A {\bf 363}, 2155
(2005).
\bibitem{eides}
M. I. Eides, H. Grotch and V. A. Shelyuto, Phys. Rept. {\bf 342}, 63
(2001).
\end{thebibliography}
\end{document}